\documentclass[preprint]{elsarticle}

\biboptions{sort&compress}

\usepackage[hypertexnames=false,breaklinks=true,colorlinks=true]{hyperref}

\newcommand{\Bernoulli}{\mathsf{Bernoulli}}
\newcommand{\Binomial}{\mathsf{Binomial}}
\newcommand{\Poisson}{\mathsf{Poisson}}
\newcommand{\Multinomial}{\mathsf{Multinomial}}
\newcommand{\Hypergeometric}{\mathsf{Hypergeometric}}
\newcommand{\Uniform}{\mathsf{Uniform}}

\usepackage{algorithm}
\usepackage{algpseudocode}

\usepackage{xspace}

\usepackage{amsmath,amsthm}
\usepackage{thmtools}
\usepackage{amsfonts,dsfont,mathtools,thm-restate}
\usepackage{amssymb}

\usepackage[capitalise, nameinlink]{cleveref} %
\crefname{ineq}{inequality}{inequalities}
\crefname{lem}{Lemma}{Lemmas}
\crefname{rmk}{Remark}{Remarks}
\crefname{equation}{}{}

\newcommand{\Ex}[1]{\mathbf{E} \left[\,#1\,\right]}
\newcommand{\Whp}{w.h.p.\xspace}
\newcommand{\Pro}[1]{\mathbf{Pr} \left[\,#1\,\right]}
\newcommand{\N}{\mathbb{N}}
\newcommand{\polylog}{\operatorname{polylog}}

\newtheorem{thm}{Theorem}
\newtheorem{lem}[thm]{Lemma}
\newtheorem{cor}[thm]{Corollary}
\newdefinition{rmk}{Remark}

\usepackage{nth}
\usepackage{ifthen}

\newboolean{UseFullJournal}
\setboolean{UseFullJournal}{false}

\newcommand{\AlgorithmicaJournal}{
\ifthenelse{\boolean{UseFullJournal}}{Algorithmica. An International Journal in Computer Science}{Algorithmica}}

\newcommand{\TwoChoice}{\textsc{Two-Choice}\xspace}
\newcommand{\TwoSample}{\textsc{Two-Sample}\xspace}
\newcommand{\Threshold}{\textsc{Threshold}\xspace}
\newcommand{\Thinning}{\textsc{Thinning}\xspace}

\usepackage{framed}

\allowdisplaybreaks

\title{An asymptotically optimal algorithm for generating bin cardinalities}

\author[1]{Luc Devroye\fnref{fn2}}
\ead{lucdevroye@gmail.com}
\affiliation[1]{organization={School of Computer Science, McGill University},
addressline={3480 Rue University},
postcode={QC H3A 2A7},
city={Montreal},
country={Canada}}

\author[2]{Dimitrios Los\corref{cor1}%
\fnref{fn2}}
\ead{mail@dimitrioslos.com}
\affiliation[2]{organization={Department of Computer Science and Technology, University of Cambridge},
addressline={15 JJ Thomson Avenue},
postcode={CB3 0FD},
city={Cambridge},
country={United Kingdom}}

\cortext[cor1]{Corresponding author}
\fntext[fn1]{Luc Devroye was supported by NSERC grant A3456.}
\fntext[fn2]{Dimitrios Los was supported by the LMS Early Career Fellowship (ref.~2024-19).}

\begin{document}

\begin{keyword}
Random variate generation \sep Simulation \sep Expected time analysis \sep Balls-into-bins \sep Random allocations \sep Hashing %
\sep \MSC[2010] 65C10 \sep 60C05 \sep 11K45 \sep 68W20 \sep 68U20 \sep 68W27 \sep 68W40
\end{keyword}

\begin{abstract}
In the balls-into-bins setting, $n$ balls are thrown uniformly at random into $n$ bins. The na\"{i}ve way to generate the final load vector takes $\Theta(n)$ time. However, it is well-known that this load vector has with high probability bin cardinalities of size $\Theta(\frac{\log n}{\log \log n})$. Here, we present an algorithm in the RAM model that generates the bin cardinalities of the final load vector in the optimal $\Theta(\frac{\log n}{\log \log n})$ time in expectation and with high probability.

Further, the algorithm that we present is still optimal for any $m \in [n, n \log n]$ balls and can also be used as a building block to efficiently simulate more involved load balancing algorithms. In particular, for the \TwoChoice algorithm, which samples two bins in each step and allocates to the least-loaded of the two, we obtain roughly a quadratic speed-up over the na\"{i}ve simulation. 

\end{abstract}

\maketitle

\section{Introduction}

In the \textit{balls-into-bins} setting, $n$ balls are allocated uniformly at random (u.a.r.) into $n$ bins. 
In this note, we present an algorithm that permits one to efficiently generate a vector of bin cardinalities. We denote by $N_i$ the number of balls ending up in bin $i$. Then, we define the vector of \textit{bin cardinalities} $(X_0, X_1, \ldots, X_n)$, where 
\[
X_j = \sum_{i = 1}^n \mathbf{1}_{\{ N_i = j \}}, \quad 0 \leq j \leq n. 
\]
Let $K_n = \max\{j : 0 \leq j \leq n, X_j > 0 \} = \max\{ N_i : 1 \leq i \leq n\}$ be the \textit{maximum occupancy}. We would like to generate $(X_0, \ldots, X_{K_n}, K_n)$ in an efficient manner. As a by-product, this would yield a way of generating $(X_{K_n}, K_n)$ which represents the number of bins with maximum occupancy, jointly with the maximum occupancy, $K_n$.

Throughout, we assume that we are in the idealized RAM model, in which real numbers can be stored and all standard operations on these numbers take constant time. This includes addition, multiplication, division, comparison, truncation, as well as exponentiation, logarithmic and trigonometric operations. We also have a generator at our disposal that produces an independent identically distributed (i.i.d.) sequence of $\Uniform(0, 1)$ random variables $U_1, U_2, \ldots $, at a unit cost per $U_i$. In this virtual model, one can generate $\Binomial(n, p)$ and $\Poisson(\lambda)$ random variables in expected time uniformly bounded over all parameters $n$, $p$ and $\lambda$. For references and algorithms, we refer to~\cite{D86}.

Just mimicking the binning process, yields a trivial linear time algorithm. However, it is known that
\[
  \frac{K_n}{\log n/\log \log n} \to 1
\]
in probability as $n \to \infty$~\cite{G81} (cf.~\cite[Theorem 4.4]{D86book} and~\cite{RS98}) and $\Ex{K_n^{\alpha}} \sim \big( \frac{\log n}{\log \log n}\big)^{\alpha}$ for all $\alpha \geq 1$~\cite[Theorem 4.4]{D86book}. Therefore, it is not unreasonable to hope for an algorithm with polylogarithmic expected time. We present an algorithm with $
  O( \log n/\log \log n)
  $
time in expectation and with high probability.\footnote{In this work, with high probability (w.h.p.) refers to probability at least $1 - o_n(1)$, where $n$ is the number of bins.} 
This is optimal as the output vector is \Whp~of size $\Omega(\log n/\log \log n)$. We note that this algorithm also has optimal performance guarantees, in the case where we throw $m$ balls into $n$ bins, for any $n \leq m \leq n \log n$. Finally, we make use of this efficient algorithm to obtain an almost quadratic speedup for the generation of bin cardinalities for more involved balanced allocation processes. This includes the \TwoChoice process, where for each ball, two bins are sampled uniformly at random and the ball is then allocated to the least loaded of the two.

\paragraph{Organization} In \cref{sec:prelims}, we introduce the notation and briefly overview related work for the probability distributions and balanced allocation processes that we use. In \cref{sec:generate_balls_and_bins}, we present the main algorithm (\cref{alg:one_choice}) for generating the bin cardinalities in optimal time (\cref{thm:main_theorem}). In \cref{sec:generate_balanced_allocations}, we use this algorithm to speed up the simulation of other balanced allocation processes (\cref{thm:two_sample}). Finally, in \cref{sec:conclusions}, we conclude with a summary and open problems. 

\section{Preliminaries} \label{sec:prelims}

\paragraph{Multinomial distribution} We say $X = (X_1, \ldots, X_k)$ is a  $\Multinomial(n; p_1, \ldots, p_k)$ distribution where $\sum_{j = 1}^k p_j = 1$, when for all non-negative integers $n_1, \ldots, n_k$ with $\sum_{j = 1}^k n_j = n$, 
\[
  \Pro{\bigcap_{j = 1}^k \left\{ X_j = n_j \right\} } = \frac{n!}{\prod_{j = 1}^k n_j!} \cdot \prod_{j = 1}^k p_j^{n_j}.
\]
The $\Multinomial(m; 1/n, \ldots, 1/n)$ distribution corresponds to balls-into-bins with $m$ balls and $n$ bins. Random variates from this distribution can be generated sequentially starting with $X_1$, which is $\Binomial(n, p_1)$. Then $X_2$ is $\Binomial(n - X_1, p_2)$, $X_3$ is $\Binomial(n - X_1 - X_2, p_3)$, and so forth. As a binomial random variate can be obtained in $O(1)$ expected time, uniformly over all choices of parameters, the multinomial can thus be obtained in $O(k)$ expected time. When $k = \infty$, then the above procedure can be stopped after $\ell$ steps, where $\ell = \min\{ j : 1 \leq j, X_1 + \ldots + X_j = n \}$, with the understanding that we return only $X_1, \ldots, X_{\ell}$. The time required for the generation is then equal to $O(\max\{ j : 1 \leq j, X_j > 0 \} )$.

\paragraph{Multivariate Hypergeometric distribution} Next, let $X = (X_1, X_2, \ldots, X_k)$ be a $ \Hypergeometric(n; m; n_1, n_2, \ldots, n_k)$ distribution where all parameters are positive integers, $\sum_{j = 1}^k n_j = n$ and $m \leq n$. This corresponds to sampling $m$ balls uniformly without replacement from an urn with $n$ balls, $n_j$ of which have the $j$-th color. For $\ell_1 \leq n_1, \ell_2 \leq n_2, \ldots$, we have that 
\[
  \Pro{\bigcap_{j = 1}^k \left\{ X_j = \ell_j \right\}} = \frac{\prod_{j = 1}^k \binom{n_j}{\ell_j} }{\binom{n}{m}}.
\]
Random variates from this distribution can also be generated sequentially, starting with $X_1$ which is $\Hypergeometric(n; m; n_1, n - n_1)$. Then $X_2$ which is $\Hypergeometric(n - n_1; m - X_1; n_2, n - n_1 - n_2)$, and so forth. As the univariate $\Hypergeometric(n; m; b, n - b)$ distribution is log-concave, and the location of its mode is known to be either at $\lceil r - 1 \rceil$ or at $\lfloor r \rfloor$, where $r = (m + 1) (b+1) / (n+2)$, we can apply a uniformly fast generator developed by Devroye~\cite{D87}. It only requires $O(1)$ time access to the value of the distribution at every point. This would require that we have an $O(1)$ time oracle at our disposal for the gamma function. However, as the rejection method is used for generating a random variate from a log-concave distribution, one can use a first-order set of Stirling bounds on the factorial, 
\[
\left( \frac{n}{e} \right)^n \cdot \sqrt{2\pi n} \cdot e^{\frac{1}{12n + 1}} \leq n! \leq \left( \frac{n}{e} \right)^n \cdot \sqrt{2\pi n} \cdot e^{\frac{1}{12n}},
\]
to accept or reject with probability close to one. If needed, we can carry out the full factorial multiplication for $n!$ in $O(n)$ time, but this is only required with probability $e^{-\Omega(n)}$, which makes the expected time to decide on exact acceptance or rejection in the rejection method $O(1)$. Some examples in this line of reasoning are given in Devroye~\cite{D86}. The entire procedure for the multivariate hypergeometric takes expected time $O(k)$ and $O(k + \ell)$ time with probability at least $1 - e^{-\Omega(\ell)}$.\footnote{In~\cite{D87}, the concentration statement is not mentioned. However, as the  algorithm is based on rejection sampling with constant acceptance probability, it follows trivially that its running time is dominated by a geometric random variable with a constant parameter, and so it has exponential tails.} The above procedure for the multivariate hypergeometric can be stopped after $\ell$ steps, where $\ell = \min\{j : 1 \leq j, X_1 + \ldots + X_j = m \}$, with the understanding that we return only $X_1, \ldots, X_{\ell}$. Then, the time required for this is equal to $O(\max\{ j : 1 \leq j, X_j  > 0 \})$.

\paragraph{Load balancing algorithms} Balls-into-bins can be used to model hashing with perfect hash functions, but it can also be used as a simple algorithm for load balancing. An improvement over this algorithm is the \TwoChoice process, which is defined as follows:

\begin{framed}
\vspace{-.45em} \noindent
\underline{\TwoChoice Process:} \\
\textsf{Iteration:} For each $t \geq 0$, sample \textit{two} bins $i_1$ and $i_2$, independently and uniformly at random. Let $i \in \{i_1, i_2 \}$ be such that $N_{i}^{t} = \min\{ N_{i_1}^t, N_{i_2}^t\}$, breaking ties randomly. Then update:  
    \begin{equation*}
     N_{i}^{t+1} = N_{i}^{t} + 1.
 \end{equation*}
 \vspace{-0.75cm}
\end{framed}

Azar, Broder, Karlin \& Upfal~\cite{ABKU99} (and Karp, Luby \& Meyer auf der Heide~\cite{KLM96}) showed that this process achieves \Whp~a maximal occupancy of $\log_2 \log n + \Theta(1)$, when $m = n$. More generally, Berenbrink, Czumaj, Steger \& V\"ocking~\cite{BCSV06} showed that it achieves \Whp~a maximal occupancy of $m/n + \log_2 \log n + \Theta(1)$, for any $m \geq n$.

In \cref{sec:generate_balanced_allocations}, we present an efficient algorithm for simulating \TwoChoice. The algorithm also applies to a wider family of processes, namely to any process that samples two bins and uses a decision function $Q$ to allocate to one of the two bins based on their loads. This should enable more thorough empirical comparisons of load balancing algorithms, which are commonplace in the theoretical load balancing literature (e.g.,~\cite[Section 7]{ABKU99},~\cite[Section 1.2]{P17},~\cite[Section 6]{ACMR98},~\cite[Section 12]{LS23Noise}).

More formally, the family of $\TwoSample$ processes is defined as follows:

\begin{samepage}
\begin{framed}
\vspace{-.45em} \noindent
\underline{$\TwoSample(Q)$ Process:} \label{sec:two_sample_def} \\
\textsf{Parameter:} a \textit{decision function} $Q: \N \times \N \to \{ 0, 1 \}$ \\
\textsf{Iteration:} For each $t \geq 0$, sample two bins $i_1$ and $i_2$, independently and uniformly at random. Then, update $N_i^{t+1} = N_i^t + 1$, where
    \begin{equation*}
     i = \begin{cases}
         i_1 & \text{if }Q(N_{i_1}^t, N_{i_2}^t) = 0, \\
         i_2 & \text{if }Q(N_{i_1}^t, N_{i_2}^t) = 1.
     \end{cases}
 \end{equation*}\vspace{-0.3cm}
\end{framed}
\end{samepage}

The \TwoChoice process is an instance of this family with $Q(N_{i_1}^t, N_{i_2}^t)= \mathbf{1}_{\{N_{i_1}^t > N_{i_2}^t\}}$. Several other well-studied processes belong to this family, such as the \Threshold processes~\cite{M99,IK04,LS22,LSS22}, where the decision function is based on a fixed threshold $f$, i.e.,
\[
  Q_f(N_{i_1}^t, N_{i_2}^t) = \begin{cases}
      0 & \text{if }N_{i_1}^t \leq f, \\
      1 & \text{otherwise}.
  \end{cases}
\]
These \Threshold processes belong to the family of \Thinning processes~\cite{FG18}, which have the several advantages of \TwoChoice, the main one being that they relax the communication requirements between the two samples.

\paragraph{Related work} The main algorithm of this work belongs to the paradigm studied by Devroye~\cite{D92} ``that various random objects defined in terms of random processes can be generated quite efficiently without `running' or `simulating' the defining process''. In that paper, the author gave efficient algorithms for generating the convex hull of a set of random points, the absorption time of a Markov chain, sums of random variables and random binary trees. Earlier examples of this paradigm include efficient generation of the maximum of $n$ i.i.d.~samples of a distribution, as in Devroye~\cite{D80}, or more generally their order statistics, as in Bentley \& Saxe~\cite{BS80} and H\"ormann \& Derflinger~\cite{HD02}.  Recently, Barak{-}Pelleg \& Berend~\cite{BpB22} introduced an efficient algorithm for sampling the coupon collector time, i.e., the time until every bin has at least one ball, that is $T = \inf \{ t : 1 \leq t, \bigcap_{i = 1}^n \{ N_i^t \geq 1 \} \}$.

For random variate generation in general, see Devroye~\cite{D86} and H\"ormann, Leydold \& Derflinger~\cite{HLD04}. For random variate generation for the $\Poisson$ distribution, see~\cite{DA74,F76,A79,AD80,D81,SK81}, for log-concave distributions, see~\cite{H94,H95,D84,D87} and for the $\Multinomial$ and $\Hypergeometric$ distributions, see~\cite{BB84,S89,S90,D93}.

For balanced allocation processes, see the books by Kolchin \& Sevastyanov~\cite{KSC78}, Johnson and Kotz~\cite{JK77}, Mahmoud~\cite{M09} and surveys by Kotz \& Balakrishnan~\cite{KB97}, Mitzenmacher, Richa \& Sitaraman~\cite{MRS01} and Wieder~\cite{W17}.

\section{Generating bin cardinalities for balls-into-bins} \label{sec:generate_balls_and_bins}

In this section, we  present the main algorithm (\cref{alg:one_choice}) for the efficient simulation of the balls-into-bins process.

\begin{thm} \label{thm:main_theorem}
For any $n \leq m \leq n \log n$, \cref{alg:one_choice} generates the vector of bin cardinalities for the $\Multinomial(m; 1/n, \ldots, 1/n)$ distribution in time 
\[
O\left( \frac{\log n}{\log \big( \frac{n}{4m} \log n \big)} \right),
\]
in expectation and with probability at least $1 - o_n(1)$.
\end{thm}

\begin{rmk}
By the lower bound on the maximum occupancy (see e.g.,~\cite{RS98} and \cite[Lemma 14]{ACMR98}), it follows that this time is asymptotically optimal.
\end{rmk}

\begin{algorithm}
\caption{\textsc{Generate-Bin-Cardinalities}($n$, $m$) \\
The algorithm for generating the bin cardinalities $X$ for $m$ balls into $n$ bins. The subroutine $\textsc{Combine-And-Sum-Cardinalities}(X,Y)$ is as outlined in \cref{cor:combine_and_sum}. The parameter $K^*$ trades-off running time and probability guarantee, see constraints in \cref{thm:aux_main_theorem}.}\label{alg:one_choice}
\begin{algorithmic}
\If{$m \leq K^*$} \Comment{Base case}
   \State {Add $m$ balls u.a.r.~to the bins to obtain $X$, using na\"ive simulation.}
   \State \Return $X$
\EndIf
\State {$N \gets 0 \qquad \vartriangleright$ Number of balls allocated}
\State {$k \gets 0\qquad \vartriangleright$ Current load}
\State {$B \gets n \qquad \vartriangleright$ Remaining bins}
\State {$\lambda \gets m - m^{3/5} \qquad \vartriangleright$ Rate of Poisson r.v.}
\State {$p \gets e^{-\lambda} \qquad \vartriangleright$ Current probability}
\State {$s \gets 0 \qquad \vartriangleright$ Current cumulative probability}
\While{$B > 0$} \Comment{Generation using Poissionization}
  \State {$X_k \gets \Binomial(B, 1/(1 - s))$}
  \State {$B \gets B - X_k$}
  \State {$N \gets N + X_k \cdot k$}
  \State {$s \gets s + p$}
  \State {$p \gets p \cdot \lambda/(k+1) $}
  \State {$k \gets k + 1$}
\EndWhile
\If{$N < m - 2m^{3/5}$} \Comment{\textbf{Case A:} We need to add many balls (\textit{rare case}).}
    \State {Add $m - N$ balls u.a.r.~to the bins in $X$ to obtain $X'$, using na\"ive simulation}
    \State {\Return $X'$}
\ElsIf{$N > m$} \Comment{\textbf{Case B:} We need to remove balls (\textit{rare case}).}
    \State {Sample $N - m$ balls u.a.r.~to remove from $X$ to obtain $X'$.} 
    \State {\Return $X'$}
\Else \Comment{\textbf{Case C:} We need to add few balls (\textit{common case}).}
    \State {$Y = \textsc{Generate-Bin-Cardinalities}(n, m - N)$} \Comment{Proceed recursively}
    \State {\Return $\textsc{Combine-And-Sum-Cardinalities}(X, Y)$}
\EndIf
\end{algorithmic}
\end{algorithm}

We start by presenting an auxiliary algorithm for efficiently combining the bin cardinalities from two different simulations with $m_x$ and $m_y$ balls respectively, in order to get the bin cardinalities for $m_x + m_y$ balls. Below we prove a more general version that we also use when simulating more involved balanced allocation processes in the next section in \cref{thm:two_sample}.

\begin{lem} \label{lem:combine_cardinalities}
Let $X$ and $Y$ be the vectors of bin cardinalities for a sample generated  from a $\Multinomial(m_x; 1/n, \ldots, 1/n)$ and a $\Multinomial(m_y; 1/n, \ldots, 1/n)$ respectively, then we can generate the bin cardinalities for the joint distribution in $O(|X| \cdot |Y|)$ expected time and $O(|X| \cdot |Y| + \ell \cdot |X|)$ time with probability at least $1 - e^{-\ell }$ (for any $\ell \geq 1$).
\end{lem}

We call the subroutine for generating a sample from the joint distribution $\textsc{Combine-Cardinalities}(X, Y)$.

\begin{proof}
We will construct the output vector $Z$, where $Z_{x,y}$ is the number of bins that were selected $x$ times in $X$ and $y$ times in $Y$. The construction is incremental where in the $r$-th step, we generate the entries $Z_{r, \cdot}$, i.e., those sampled $r$ times in $X$, while keeping track of $T$, the bins not yet processed in $Y$. So, initially $T = Y$.

In the $r$-th step, we sample without replacement $X_r$ bins from the bins in $T$, which follows the multivariate $Z_{r, \cdot} \sim \Hypergeometric(\sum_i T_i; X_r; T_0, \ldots , T_{|Y|-1})$ distribution. Then, we append these to $Z$ and remove the respective counts from $T$. The time to complete each step is dominated by a geometric random variable with expectation $O(|Y|)$, and so overall this algorithm requires $O(|X| \cdot |Y|)$ expected time and has exponential tails, i.e., for any $\ell \geq 1$ it takes $O(|X| \cdot |Y| + \ell \cdot |X|)$ time with probability at least $1 - e^{-\ell}$ (for any $\ell \geq 1$).
\end{proof}

By aggregating the entries in the joint distribution with the same sum, we obtain the following corollary.

\begin{cor} \label{cor:combine_and_sum}
Let $X$ and $Y$ be the vectors of bin cardinalities for a sample generated from $\Multinomial(m_x; 1/n, \ldots, 1/n)$ and $\Multinomial(m_y; 1/n, \ldots, 1/n)$ independently, then we can combine them to generate the bin cardinalities of a sample from $\Multinomial(m_x + m_y; 1/n, \ldots, 1/n)$ in $O(|X| \cdot |Y|)$ expected time and $O(|X| \cdot |Y| + \ell \cdot |X|)$ time with probability at least $1 - e^{-\ell }$ (for any $\ell \geq 1$).
\end{cor}

We call this algorithm $\textsc{Combine-And-Sum-Cardinalities}(X, Y)$ and we will use it in the proof of the main theorem below. 

We are going to prove a slightly stronger version of \cref{thm:main_theorem}, which allows us to obtain a tradeoff between the running time of the algorithm and its probability guarantee. This will be useful in \cref{sec:generate_balanced_allocations} where we need a $\polylog(n)$ time algorithm with probability $1 - n^{-\Omega(1)}$.

\begin{thm} \label{thm:aux_main_theorem}
Consider any $n \leq m \leq n \log n$ and any $K^*$ such that
\begin{align*} 
10 \cdot \frac{\log n}{\log \left(\frac{4n}{m} \log n \right)} \leq K^* \leq 8 \cdot (\log n)^5.
\end{align*}
Then, \cref{alg:one_choice} with parameter $K^*$ generates the vector of bin cardinalities for the $\Multinomial(m; 1/n, \ldots, 1/n)$ distribution in time $O(K^*)$ in expectation and with probability at least $1 - e^{-\frac{1}{4}(K^*)^{1/5}}$.
\end{thm}

The main theorem (\cref{thm:main_theorem}) follows for $K^* = 10 \cdot \frac{\log n}{\log \left(\frac{4n}{m} \log n \right)}$.

\begin{proof}[Proof of \cref{thm:aux_main_theorem}]
Let us throw $N$ balls uniformly and independently into $n$ bins, where $N$ is $\Poisson(\lambda)$. Then, the bin sizes $N_i$ are independent $\Poisson(\lambda/n)$ random variables (cf.~\cite[Chapter 5.1]{MU17}). This implies that $(X_0, X_1, \ldots )$ is $\Multinomial(n; p_0, p_1, \ldots )$, where $p_i$ is the probability that one bin has size $i$, i.e.,
\[
  p_i = \Pro{\Poisson(\lambda/n) = i}, \quad i \geq 0.
\]
The $p_i$'s can be computed recursively and on the fly and we will never need more than $K = \max\{ i : 1 \leq 1, X_i >0 \}$ of them. This is distributed as the maximum of $n$ independent $\Poisson(\lambda/n)$ random variables, which, for $\lambda = O(n)$, is bounded by $(1+o(1)) \cdot \frac{\log n}{\log \log n}$. Thus, the $\Multinomial$ random vector $(X_0, X_1, \ldots, X_K)$ can be generated in expected time $O(K)$, which is $O(\frac{\log n}{\log \log n})$. It is understood that $X_{K + i} = 0$ for all $i > 0$.

If we take $\lambda = m$, then the number of balls, $N$, is close to, but not equal to $m$. A small adjustment is required, which we present below. As the difference $|N - m|$ is in expectation $\Theta(\sqrt{m})$, it is too large to adjust point by point, repeatedly adding or removing one randomly picked element. Such a procedure would lead to an additional expected time complexity of the order of $\Theta(\sqrt{m})$, which is unacceptable. 

From the Poissonized sample with $\lambda = m$, we obtain a random vector $(X_0, X_1, \ldots, X_K)$, where $K$ is distributed as the maximum of $n$ independent $\Poisson(m/n)$ random variables. Since $\sum_{i = 1}^K iX_i = N$, where $N$ is $\Poisson(m)$, an adjustment may be required to correct the sample size when $N \neq m$. If there is a surplus, i.e., $N > m$, then we need to remove $N - m$ different randomly picked elements. This, however, is rather cumbersome, while adding elements is much more straightforward. 

To fix this conundrum, we strategically pick $\lambda = m - m^{3/5}$.\footnote{Actually, we could have chosen $\sqrt{3 m \log m}$ instead of $m^{3/5}$, and still obtain the fast running time with probability at least $1 - o(1)$. But in \cref{thm:two_sample}, we will need probability at least $1 - n^{-\Omega(1)}$, which we obtain for a slightly different value of $K^*$.} The likelihood of a surplus is now much reduced, so that it is easy to remove the $N - m$ excess elements individually. First of all,
\begin{align*}
  \Ex{(N - m)_+} 
    & = \sum_{j = m+1}^{\infty} j \cdot \frac{\lambda^j \cdot e^{-\lambda}}{j!} \\
    & = \lambda \cdot \sum_{j = m}^{\infty} \frac{\lambda^j \cdot e^{-\lambda}}{j!} \\
    & = \lambda \cdot \Pro{\Poisson(\lambda) \geq m} \\
    & \leq \lambda \cdot e^{m - \lambda - m \log(m/\lambda)} \text{ (by Chernoff's bound)} \\
    & = \lambda \cdot e^{m^{3/5} + m \log \left( 1 - \frac{m^{3/5}}{m} \right)} \\
    & \leq m \cdot e^{-m^{1/5}} \\
    & < e^{-\frac{1}{2}m^{1/5}},
\end{align*}
using Chernoff's bound~\cite{C52} (cf.~\cite{BLM13}), the inequality $\log(1 - x) \leq -x - x^2/2$ for any $x \in (0, 1)$ and that $\lambda \leq m$. By Markov's inequality, we also have that
\begin{align} \label{eq:case_b_prob_bound}
  \Pro{(N - m)_+ < 1} = %
  1 - \Pro{(N - m)_+ \geq 1}
  \geq 1 - e^{-\frac{1}{2}m^{1/5}}.
\end{align}
Similarly, we obtain that $(m - 2m^{3/5} - N)_+ < e^{-\frac{1}{2}m^{1/5}}$ in expectation and
\begin{align} \label{eq:case_a_prob_bound}
  \Pro{(m - 2m^{3/5} - N)_+ < 1} 
  \geq 1 - e^{-\frac{1}{2}m^{1/5}}.
\end{align}

Recall that \begin{align} \label{eq:k_star}
K^* \geq 10 \cdot \frac{\log n}{\log \left(\frac{4n}{m} \log n \right)},
\end{align}
which bounds the maximum occupancy in expectation and with probability at least $1 - n^{-3}$ for any $n \leq m \leq n \log n$ (see e.g.,~\cite[Lemma 14]{ACMR98}).

\textbf{Case A [$N < m - 2m^{3/5}$]:} If  the deficit $m - N$ is more than $2m^{3/5}$, then we add items one by one until the deficit is $0$. Each addition is done first by picking a random $i$ with probability $X_i/n$, decreasing $X_i$ by one, and increasing $X_{i+1}$ by one. Each such selection takes expected time not exceeding $O(K^*)$. Therefore the total expected time for reducing the deficit to $0$ is bounded by\[
m^{3/5} \cdot \Ex{(m - 2m^{3/5} - N)_+} \cdot O(K^*)
 = m^{3/5} \cdot e^{-\frac{1}{2} m^{1/5}} \cdot O(K^*) 
 = O(K^*).
\]
Further, by \cref{eq:case_a_prob_bound}, \Whp this case is not encountered.

\medskip

\textbf{Case B [$N > m$]:} If on the other hand there is an excess, we remove an excess item by picking an integer $i$ with probability $i X_i /N$, as this identifies a uniformly random item in a bin with cardinality $i$. This can be done in time not exceeding the number of choices, which in this case is in expected value (and \Whp) $O(K^*)$. Having chosen such an $i$, we decrease $X_i$ (and thus $N$) by one. This is repeated until $N$ reaches $m$. The total expected work for the removal procedure is thus bounded by \[
  \Ex{(N- m)_+} \cdot O\left( K^* \right)= O\left( K^* \right).
\]
Again, by \cref{eq:case_b_prob_bound}, \Whp this case is not encountered.

\textbf{Case C [$m - 2m^{3/5} \leq N \leq  m$]:} For this case, we proceed by recursively calling the algorithm with parameters $n' = n$ and $m' = m - N$, obtaining the vector $Y = (Y_0, \ldots, Y_M)$. Note that $M \leq 10$ with high probability, since
\[
  \Pro{M \geq 11} \leq \binom{m'}{11} \cdot \left(  \frac{1}{n} \right)^{10} = O\left( \frac{(m')^{11}}{n^{10}} \right) = O\left( \frac{1}{n^{3}} \right),
\]
using that $m' \leq 2m^{3/5} \leq 2(n \log n)^{3/5}$. Also, it holds that
\begin{align*}
  \Ex{M} & = \sum_{j = 1}^{\infty} \Pro{M \geq j} \\
    & \leq 4 + \sum_{j = 5}^{\infty} \binom{m'}{j} \cdot \left( \frac{1}{n} \right)^{j-1} \\
    & = 4 + O(n) \cdot \sum_{j=5}^{\infty} \left( \frac{1}{n^{3}} \right)^j \\
    & = 4 + o(1).
\end{align*}
Using the \textsc{Combine-And-Sum} procedure in~\cref{cor:combine_and_sum} we can combine $X$ with $Y$ in $O(|X| \cdot |Y|)$ expected time and $O(|X| \cdot |Y| + \ell \cdot |Y|)$ with probability at least $1 - e^{-\ell}$, for some $\ell \geq 1$ to be chosen below.

We stop the recursion once there are $K^*$ balls left and then we na\"ively simulate the remaining balls in $O(K^*)$ time. 

Let $r$ be the number of recursive calls and  $m_1, \ldots, m_r$ the number of balls allocated in each of these. Since $m_{i+1} \leq 2m_i^{3/5}$, we have that $r = O(\log \log n)$. Further, let $T_1, \ldots, T_r$ be the running time for each recursive call. The first call takes $\Ex{T_1} = O(K^*)$, the last call takes $T_k = O(K^*)$ and all others take $\Ex{T_i} = O(1)$ time. So, since $K^* \geq r$, overall the algorithm takes $O(K^*)$ time in expectation.

Now we turn to proving the high probability bound. Let $\mathcal{B}_i$ be the event that in the $i$-th call either of the following \textit{bad events} occur: $(i)$ the maximum occupancy is $> K^*$, $(ii)$ we encounter Case A or Case B or $(iii)$ the set $Y$ has $|Y| \geq 11$. We upper bound the probability of any of these events occurring as follows:
\begin{align*}
\Pro{\bigcup_{i \in [r]} \mathcal{B}_i} 
 & \leq (n^{-3} + n^{-3} + 2 \cdot e^{-\frac{1}{2} m^{1/5}}) \cdot O(\log \log n) \\
 &  \leq 3 \cdot e^{-\frac{1}{2} m^{1/5}} \cdot O(\log \log n).
\end{align*}
By conditioning on none of the bad events occurring, the running times of the \textsc{Combine-And-Sum} procedure is dominated by a sum of independent geometric random variables. Hence, using a Chernoff bound for $\ell = \frac{1}{2} (K^*)^{1/5}$, their aggregate running time is $O(K^*)$ (constant factor times the expectation) with probability at least $1 - e^{-\frac{1}{2}(K^*)^{1/5}}$. By combining these two events, we get that the total running time is $O(K^*)$ with probability at least
\[
 \left( 1 - 3 \cdot e^{-\frac{1}{2} m^{1/5}} \cdot O(\log \log n) \right) \cdot \left(1 - e^{-\frac{1}{2}(K^*)^{1/5}} \right) \geq 1 - e^{-\frac{1}{4} (K^*)^{1/5}},
\]
using that $m \geq K^*$.
\end{proof}

\section{Generating bin cardinalities for \TwoSample processes} \label{sec:generate_balanced_allocations}

In this section, we will use \cref{alg:one_choice} to speed up the simulation time of any \TwoSample process, defined in \cref{sec:two_sample_def}, from $\Theta(m)$ time down to $\tilde{O}(\sqrt{n})$ for any $n \leq m \leq n \log n$.

\begin{thm} \label{thm:two_sample}
For any $\TwoSample(Q)$ process where the decision function $Q$ can be computed in $O(1)$ time, we can simulate the execution of the process for $n \leq m \leq n \log n$ balls in $O(\sqrt{n} \cdot (\log n)^6 )$ time in expectation and \Whp
\end{thm}

On a high level, the proposed algorithm splits the allocations in blocks of $\Theta(\sqrt{n})$ consecutive allocations. In any such block, very few bins are sampled more than once and so most allocations can be simulated in batches. The key idea is to use \cref{alg:one_choice} to compute how many pairs of bins are sampled between any two load values $\ell_1$ and $\ell_2$, and then perform all these allocations in one go, as the allocated bin depends only on the values of $\ell_1$ and $\ell_2$. Finally, the allocations involving bins that were sampled more than once in a block can be naively simulated.

\begin{proof}   
Note that for any \TwoSample process, the first and second samples are generated by a balls-into-bins process, so \Whp~the maximum occupancy is $2 \cdot O(\log n)$. When sampling from this balls-into-bins process we will use \cref{alg:one_choice} with $K^* = 8 \cdot (\log n)^5$, so that by \cref{thm:aux_main_theorem} the algorithm terminates in $O( (\log n)^5 )$ time with probability at least $1 - n^{-2}$.

We will split the allocations into blocks of $M = \frac{1}{4} \sqrt{n}$ balls each. The main idea is that within each block there will be a small number of bins that are sampled more than once, and so very few bins will be allocated more than one ball. 

To verify this, let $C_i$ be the indicator of the event that any of the two samples in the $i$-th step is the same as another sample in the block. Then, 
\[
  \Pro{C_i = 1} \leq \frac{4M}{n} = \frac{1}{\sqrt{n}},
\]
and so the number of collisions $C = \sum C_i$ in the block satisfies
\[
  \Ex{C} \leq 2M \cdot \frac{1}{\sqrt{n}} = \frac{1}{2}.
\]
Further, let $C_i'$ be independent $\Bernoulli(1/\sqrt{n})$ random variables and $C' = \sum C_i'$. Then, $C'$ stochastically dominates $C$ and so using Chernoff's bound~\cite{C52} (cf.~\cite{BLM13}),
\[
  \Pro{C > 9 \log n} \leq \Pro{C' > 9 \log n} \leq n^{-2}.
\]
Therefore, by taking the union bound we have that \Whp~all blocks have at most $9 \log n$ collisions (internally).

Now for each block $(t, t + M]$, we perform the following steps:
\begin{itemize}
  \item (\textbf{Generate $Z_{\ell, x, y}$}) We generate the vector $Z$, where $Z_{\ell, x, y}$ counts the  bins with load $\ell$ at step $t$, that were sampled $x$ times as a first sample and $y$ times as a second sample in the block. 
  
  We use the hypergeometric distribution to first generate  $F_{\ell}$, the count of the number of bins with initial load $\ell$ that appeared as a first sample in the block. Then, using \cref{alg:one_choice} in $O((\log n)^5)$ time for each load value, we generate $F_{\ell, x}$ the number of bins with load $\ell$ that appeared $x$ times as a first sample. Similarly, we generate $S_{\ell, y}$ the number of bins with load $\ell$ that appeared $y$ times as a second sample. Then, we combine these two using the subroutine \textsc{Combine-Cardinalities} in \cref{lem:combine_cardinalities}, to get $Z$. Since there are $O(\log n)$ different load values and combining takes $O(|F| \cdot |S|)$ time, this step takes in total \Whp~$O( (\log n)^5)$ time.
  
  \item (\textbf{Pairing}) Next, we generate the pairs of samples for the allocations. For each of the special  $O(\log n)$ bins which have been sampled more than once in the block, i.e., $\ell$'s in $Z_{\ell, x, y}$ with $x > 1$ or $y > 1$, we assign it a special bin id. Next, we generate its pairs by sampling the hypergeometric distribution, and then adjust the $Z$ counts appropriately, giving an id to each of its paired bin (if it does not already have one). This takes $O( (\log n)^5 )$ time and generates a set of pairs $P$.
  
  Next, we pair the remaining items by sampling from a $\Hypergeometric$ distribution and again decreasing the counts of the sampled load values.  This generates a vector $P'$, where $P_{\ell_1, \ell_2}'$ gives the count of the pairs where the first sample has load value $\ell_1$ and the second load value $\ell_2$ at step $t$. Again this takes $O( (\log n)^5 )$ time.
  \item (\textbf{Simulation}) Finally, we need to simulate the allocations. We simulate $P$ and $P'$ separately as there is no overlap between their bins:
  
  For $P$, sample a random order and na\"{i}vely simulate the allocations using the decision function $Q$. Since there are $O(\log n)$ entries in $P$, this takes $O(\log n)$ time.

  Next, for each pair of load values $\ell_1, \ell_2$ with $P_{\ell_1, \ell_2}' > 0$, we know that the decision function is going to allocate to the first bin if $Q(\ell_1, \ell_2) = 0$, and otherwise to $\ell_2$. So, if $Q(\ell_1, \ell_2) = 0$, then there are going to be $P_{\ell_1, \ell_2}'$ bins that change their load from $\ell_1$ to $\ell_1 + 1$ and otherwise $P_{\ell_1, \ell_2}'$ bins that change their load from $\ell_2$ to $\ell_2 + 1$; so we adjust the counts accordingly. This takes $O((\log n)^2)$ time.

  In the end, in $O( \log n )$ time we aggregate the bins of the same resulting load values (using a hash table) and proceed to the next block.
\end{itemize}

Finally, we take the union bound over the $O(\sqrt{n} \cdot \log n)$ blocks having at most $O(\log n)$ collisions each and all executions of the modified \cref{alg:one_choice} terminating in $O((\log n)^5)$ time. So, we conclude that each block takes $O( (\log n)^5)$ time to process and the total simulation takes \Whp~$O(m/\sqrt{n} \cdot (\log n)^6) = O(\sqrt{n} \cdot (\log n)^6)$ time.
\end{proof}

The following speedup for a special case of \Thinning, such as the processes studied in~\cite{FG18}, was remarked by one of the referees.

\begin{rmk} \label{rem:thinning_with_bin_counts}
Consider the special case of \Thinning processes where thresholding is based on the number of times a particular bin was selected as the first sample. It is possible to simulate these processes using two rounds of balls-into-bins (i.e., in twice the time given by \cref{thm:main_theorem}), where in the first round all the first samples are selected and in the second round all balls above the given threshold are re-allocated randomly.
\end{rmk}

\section{Conclusions} \label{sec:conclusions}

In this note, we presented an algorithm in the RAM model for simulating the balls-into bins process in optimal $O(\log n /\log \log n)$ time in expectation and with high probability. The algorithm also applies to the setting with $m \in [n, n \log n]$ balls, giving optimal performance guarantees. Further, we used this algorithm to obtain a quadratic improvement in the time required to simulate a large family of load balancing processes, which includes the well-known \TwoChoice process.  

There are several questions that remain open. For instance, it would be interesting to extend the current algorithm for any $m > n \log n$ aiming for a time complexity of $O(\sqrt{\frac{m}{n} \cdot \log n})$, and investigate generalizations in models that are less powerful than the RAM model. 

Further, it is natural to ask whether the \TwoChoice process can be simulated in $\polylog(n)$ time, given that \Whp~the different load values are exponentially fewer (for $m = n$) than the ones in the balls-into-bins process. Following \cref{rem:thinning_with_bin_counts}, it would be interesting to investigate the time to simulate other processes in the \Thinning family.

\section*{Acknowledgments}

\noindent The authors would like to thank both referees for their wonderful suggestions.

\paragraph{Declarations of interest} None.

\paragraph{CRediT authorship contribution statement} \textbf{Luc Devroye:} Conceptualization, Methodology, Formal analysis, Writing - Review \& Editing  \textbf{Dimitrios Los:} Conceptualization, Methodology, Formal analysis, Writing - Review \& Editing
 
\bibliography{bibliography}
\bibliographystyle{elsarticle-harv.bst}

\end{document}